# Adaptive Multi-Strategy Market-Making Agent For Volatile Markets

Ali Raheman[1], Anton Kolonin[1,2,3][0000-0003-4180-2870], Alexey Glushchenko[1][0000-0002-8183-9208], Arseniy Fokin[1][0000-0002-7868-6482], Ikram Ansari[1][0000-0002-9091-6674]

[1] Autonio Foundation Ltd., Bristol, UK
[2] SingularityNET Foundation, Amsterdam, Netherlands
[3] Novosibirsk State University, Novosibirsk, Russian Federation
{ali.raheman,akolonin}@gmail.com

**Abstract.** Crypto-currency market uncertainty drives the need to find adaptive solutions to maximize gain or at least to avoid loss throughout the periods of trading activity. Given the high dimensionality and complexity of the state-action space in this domain, it can be treated as a "Narrow AGI" problem with the scope of goals and environments bound to financial markets. Adaptive Multi-Strategy Agent approach for market-making introduces a new solution to maximize positive "alpha" in long-term handling limit order book (LOB) positions by using multiple sub-agents implementing different strategies with a dynamic selection of these agents based on changing market conditions. AMSA provides no specific strategy of its own while being responsible for segmenting the periods of market-making activity into smaller execution sub-periods, performing internal backtesting on historical data on each of the sub-periods, doing sub-agent performance evaluation and re-selection of them at the end of each sub-period, and collecting returns and losses incrementally. With this approach, the return becomes a function of hyperparameters such as market data granularity (refresh rate), the execution sub-period duration, number of active sub-agents, and their individual strategies. Sub-agent selection for the next trading sub-period is made based on return/loss and alpha values obtained during internal backtesting as well as real trading. Experiments with the AMSA have been performed under different market conditions relying on historical data and proved a high probability of positive alpha throughout the periods of trading activity in the case of properly selected hyperparameters.

## 1 Introduction

The extension of the algorithmic trading approach to the market making problem has been in the focus of the research community over the last few decades. Avellaneda has used simulation modeling for exploring how the different parameters can affect performance of the active portfolio management by means of market making operations with limit order book [Avellaneda, 2008]. The specifics of the market making risk management is related to the need of locking funds in the limit order which might be causing what is called "impermanent loss" in finance. The risks associated with inventory management in market making have been studied by Guéant [Guéant, 2012].

On the other hand, the use of machine learning algorithms such as deep learning and reinforcement learning have been actively explored in the last decade. Deep learning has been explored for price prediction by exploiting stationary limit order book features used in market making [Tsantekidis 2018]. Reinforcement learning has been tried in order to get applied for trading on conventional financial markets (futures) [Zhang 2019] as well as for market making [Ganesh, 2019]. The latest trend in using of machine learning on financial markets can be seen as an attempt to operate at the strategy level, trying to figure out the more appropriate strategies for specific market conditions instead of trying to generate sparse "trading signals", which is not that helpful when dealing with funds locked in the limit order book orders [Yanjun, 2020].

Moreover, since the boom of crypto-currency markets five years ago, attempts have been made to apply algorithmic trading powered by machine learning to market making on centralized crypto-currency exchanges such as Binance. The most interesting series of studies have been run by Sadighian in 2019-2020. His works [Sadighian, 2019] and [Sadighian, 2020] explore the possibility of using deep reinforcement learning to learn how to manage positions on the limit order book based on feedback evaluated in terms of profits and losses. Unfortunately, the results have not shown an ability to provide significant and reliable profits.

In this work, we have tried to implement the principles of "purposeful activity" [Vityaev, 2015] and "experiential learning" [Kolonin, 2021] as a "Narrow Artificial General Intelligence" (Narrow AGI) solution applied to financial active portfolio management domain. We follow the concept of an agent constantly building and updating its model of the surrounding environment, as well as trying to use this model in order to evaluate different behavioral strategies relying on episodic memories, applying the "hypothetically winning" strategies to the real operational environments as it has been attempted in [Raheman, AGI 2021] relying on simulation and backtesting framework presented in [Raheman, KNOTH 2021].

## 2 Adaptive Multi-Strategy Agent

The Adaptive Multi-Strategy Agent (AMSA) for market making approach anticipates that no reliable prediction of the market price can be made at all, due to the volatile nature of the crypto markets. That might be one of the explanations as to why no "successful stories" are attributed to attempts to apply machine learning to the market making in crypto finance. In turn, the architecture suggested in current work ensures adaptability of an agent of algorithmic market making to ongoing stochastic changes of the price as well as overall market conditions with different trends ("bear", "bull", "flat") overlapped with different levels of volatility ("high", "low") in an unpredictable manner. The concept introduced in [Raheman, AGI 2021] provides no strategy of its own using a pre-configured set of multiple agents with individual strategies instead. From the exchange perspective, it may behave like a single "chair" agent, employing "macro-strategy" while the latter might be executed in total by multiple "micro-strategies" run by its "subordinate" agents.

## 2.1 Key Principles

The suggested market-making architecture is intended for autonomous operations on the financial crypto-market. It is expected to perform purposeful activity: maximizing profits and minimizing losses given the current market conditions. This is measured as profits and losses recognized for agents running specific strategies as points in the multi-dimensional space of possible strategies where dimensions are the parameters such as bid/ask spread or limit order cancellation policy. At the same time, it refers to the historical data, involving all trades made on the particular trading pair in the recent past, as well as snapshots of the limit order book for the pair, using it to evaluate all imaginable strategies safely in the "virtual" backtesting environment as discussed in [Raheman, KNOTH 2021].

AMSA main features are:
- Implementing no strategy of its own.
- Providing real trading (real exchange) and backtesting (on historical market data) environment for child agents.
- Making all limit orders issued by "subordinate" agents on behalf of the "chair" AMSA agent.
- Keeping track of all orders made by the "subordinate" agents, evaluating their performances.
- Performing time management by splitting the market making period into execution/evaluation periods.
- Calculate total profits/losses for each period.

## 2.2 Implementation Details

There are two sets, or pools of agents, used by AMSA at the same time. One set operates with real money and another one is used for strategy evaluation on historical data (backtesting). AMSA implicitly uses three environments: a) for real market making, which is used to run all agents selected for current execution period; b) for idle agents who are not allowed to operate for real; c) one for internal backtesting containing clones of all agents in a) and b), running their strategies in a "virtual environment". Operations in real market making, backtesting, and virtual environments are performed simultaneously. Agents residing in the idle environment are not performing operations. Agents are moved between real market making (a) and idle (b) environments during assessment of their performances and selection at the end of each execution period. All agents are always involved in internal backtesting while only the best ones are involved in real trading.

## 2.3 Agent/Strategy Assessment and Selection

The selection or omission of an agent applies to the "micro-strategy" being implemented by the agent. Initial agent selection is made based on backtesting results ap-plied to the historical interval one execution period long prior to the starting time of the first execution period. For all the subsequent periods, the evaluation is done on both real market making and backtesting environments. Agents showing positive return and "alpha" (i.e., excess return compared to the buy-and-hold "hodler" strategy, which means buying base asset at the beginning of the experiment and selling it at the end) due to their strategies are selected for the next period of real market making. The real market making period is skipped for those agents which do not satisfy agent selection policy. All agents are always used for backtesting.

### 2.4 Algorithm

The following algorithm is employed to run by the AMSA on every refresh (corresponding to simulation cycle interval) in place of the conventional order handling procedure and is just running and evaluating agents of different families implementing specific strategies as described further.

**Algorithm 1** AMSA Algorithm

**Input**: Time, Price
**Parameters:** start_time, end_time, period, agents, real_env, backtest_env, inventory_history
**Output:** inventory_history
1: **if** time == start_time **then** # start of experiment
2: backtest(start_time-period, start_time) # initial
3:     period_start = start_time
4:     period_end = start_time + period_len
5: **if** time % period == 0 **then** # end of period
6:     real_inventory = count_totals(real_env)
7:     backtest_inventory = count_totals(backtest_env)
8:     best_agents = select_best(backtest_totals)
9:     real_env.agents = best_agents
10:    period_start = period_end
11: **for** agent **in** real_env.agents **do** # real market making
12:    agent.handle_orders() # create/cancel orders
13: **for** agent **in** backtest_env.agents **do** # back-testing
14:    agent.handle_orders() # create/cancel orders
15: **return** inventory_history

### 2.5 Inventory Sharing Policy

Initial inventory amounts are evenly shared between all agents for both backtesting and real market making. For subsequent periods, current inventory is evenly shared between all agents selected for real market-making while for backtesting total inventory amount stays the same equal to the initial amount, evenly shared between all agents.

## 3 Experimental Setup

### 3.1 Evaluation Environment

The evaluation environment for our experiments was backtesting by means of simulation of the conventional limit order book execution policy and relying on historical trading data including both raw trades and snapshots of the LOB, while including 50 levels of bid orders and 50 levels of ask orders as well as Open-High-Low-Close-Volume (OHLCV) frames. All snapshots and frames were available with time granularities of 1 minute and 1 hour, corresponding to respective simulation intervals. The backtesting framework described in [Raheman, KNOTH 2021] and [Raheman, AGI 2021] was simulating Binance limit order book execution policy against the historical trades based on each of the simulation intervals. In turn, the order book simulation was involving modification of the historical order book snapshots with the limit orders created by the agents involved in the simulation, so some of the historical trades were executed against actual historical LOB positions while others were "intercepted" by the "injected" positions owned by market making agents involved in simulation.

### 3.2 Three Types of Historical Market Intervals

We were running the experiments on historical market data on BTC/USDT trading pair available from `binance.com` and `cryptotick.com` with time granularities corresponding to target experimental simulation intervals (1 hour, 1 minute). Three BTC/USDT historical periods of different types of market conditions were chosen for AMSA test runs: "bull" low volatile in October 2020, "bull" highly volatile in January 2021 and "bear" in May 2021.

### 3.3 Three Sets of Market Making Agents and Hodler

Three families of market making agents were selected to be AMSA working force in the experiments: Base agents implementing basic strategies, NIOX agents, and Hummingbot agents. A collection of agents belonging to a given family being controlled by a "chair" AMSA agent might be thought of as a regular/irregular bid/ask "order grid" (may be called "staggered orders") with selective creation/cancellation of the orders on respective price levels of the limit order book. For each of the experiments, "chair" AMSA agent was credited 0.1 BTC plus the same amount in USDT according to the market price at the beginning of the experiment. These amounts have been evenly distributed across inventories of the "subordinate" agents.

"**Base**" **agents** used in our experiments are described in the earlier work [Raheman, AGI 2021]. Agents of this family may have only one limit order at a time on either the ask or bid side of the spread. A new order is created only once the current open limit order is filled. Base agents configuration may differ in bid/ask spread (five ranges) and order cancellation policies (never, always, opposite). The "never" policy means that the limit order is never cancelled until it is completely filled. The "always" policy means that the existing order is always cancelled on every agent refresh time (1 minute or 1 hour). The "opposite" policy means the current order is always cancelled when the price move changes direction. Bid/ask spreads are symmetric, so percent of the spread is the same for bid and ask orders. 27 configurations were used in our experiments in total.

**NIOX agents** were implemented as part of closed-source project reproducing the market making strategy described at `https://autonio.gitbook.io/autonio-foundation/niox-suite/maker`. The strategies of these agents were different only in bid/ask spreads (asymmetric or skewed bid/ask spread), as the previously ran experiments have shown this parameter has turned to be the key drive for profits and losses for agents of this kind under different market conditions. 50 agent configurations were used in total.

**Hummingbot agents** were implemented as a closed-source clone of the open-source Hummingbot "Pure Market Making" strategy, adapted to deal with the simulation and backtesting environment. Source code of the Hummingbot is publicly available at `https://github.com/hummingbot/hummingbot`. The Hummingbot agent grid had 6 by 6 bid/ask levels (0.3, 0.5, 0.8, 1.3, 3.4, 5.5). 36 agent configurations were used in total.

**Hodler agent** as an extra single configuration implementing the "hodler" ("buy and hold") strategy was used in each of the experiments for reference. This strategy involved just buying as much as possible of base currency at the beginning of the simulation and selling it in the end.

### 3.4 Experimental Configurations

There were multiple experimental setups ran for each of the three respective market types (bull low volatile, bull highly volatile and bear), for each of the three families of agents, for different time granularities (1 minute and 1 hour) and for four different durations of execution/evaluation periods (1, 2, 3 and 5 days), as shown on Figure 1.

## 4 Experimental results

### 4.1 Performance Comparison by Interval

The results were evaluated by assessing the return of investments (ROI), as shown for the case of 1-minute based simulation (backtesting) interval in Figure 1 on the next page. Regardless of period, all three agent families have shown positive alpha in case of bear market. Bull highly volatile market brings negative return and alpha for Base and NIOX agents while Hummingbot stays positive regardless of period. Bull non-volatile market results are highly dependent on period (growing with the period durability) for Hummingbot, being constantly negative for Base and NIOX.

In case of 1-hour based simulation Base agent configuration rarely shows positive alpha and appears highly dependent on period duration regardless of type of market. NIOX shows positive alpha for bear market with slight period dependency while remains negative for bull non-volatile market, on bear highly volatile market its alpha grows with the period duration. Hummingbot shows positive alpha for most periods in all markets, mostly successful for volatile market.

## 4.2 Performance Comparison by Market Making Agent

**Bull Non-Volatile Market**

| Period | Hodler ROI, % | | Base Makers ROI, % | | NIOX Makers ROI, % | | Hummingbot Makers ROI, % | |
|---|---|---|---|---|---|---|---|---|
| | Hours | Minutes | Hours | Minutes | Hours | Minutes | Hours | Minutes |
| 1 | 12.90 | 12.90 | 4.67 | -8.14 | -26.45 | -24.07 | 12.55 | -3.46 |
| 2 | 12.90 | 12.90 | 16.49 | -5.70 | -25.90 | -25.65 | 21.61 | 4.99 |
| 3 | 12.90 | 12.90 | -3.08 | -14.29 | -27.02 | -27.36 | 7.55 | 11.19 |
| 5 | 12.90 | 12.90 | 2.06 | -5.53 | -24.71 | -21.73 | 11.33 | 13.61 |

**Bull Highly-Volatile Market**

| Period | Hodler ROI, % | | Base Makers ROI, % | | NIOX Makers ROI, % | | Hummingbot Makers ROI, % | |
|---|---|---|---|---|---|---|---|---|
| | Hours | Minutes | Hours | Minutes | Hours | Minutes | Hours | Minutes |
| 1 | 9.20 | 9.20 | 37.14 | -21.60 | -12.88 | -16.47 | 57.09 | 27.36 |
| 2 | 9.20 | 9.20 | -7.35 | -16.99 | -7.31 | 13.67 | 26.83 | 18.75 |
| 3 | 9.20 | 9.20 | -5.51 | -59.47 | 17.89 | 10.51 | 42.62 | 18.86 |
| 5 | 9.20 | 9.20 | 42.82 | -17.65 | 58.16 | 5.09 | 76.80 | 57.80 |

**Bear Market**

| Period | Hodler ROI, % | | Base Makers ROI, % | | NIOX Makers ROI, % | | Hummingbot Makers ROI, % | |
|---|---|---|---|---|---|---|---|---|
| | Hours | Minutes | Hours | Minutes | Hours | Minutes | Hours | Minutes |
| 1 | -19.10 | -19.10 | 2.45 | 3.03 | 34.23 | 35.06 | 6.79 | 26.46 |
| 2 | -19.10 | -19.10 | -5.95 | 23.20 | 35.99 | 34.60 | 27.07 | 40.11 |
| 3 | -19.10 | -19.10 | 8.71 | 52.19 | 41.20 | 48.69 | 35.88 | 45.11 |
| 5 | -19.10 | -19.10 | 6.99 | 66.16 | 34.48 | 46.63 | 48.59 | 51.52 |

**Base** makers are consistently effective on bear market, showing much better result on minutely data. Only 2-day period on minute data has positive alpha for bull non-volatile market. Bull volatile market is a complete loss on minutes while depends on period duration in case of hours.

**NIOX** is constantly losing on bull non-volatile market, unstable on bull volatile market and has a good performance on bear market for both hours and minutes.

**Hummingbot** has constantly positive alpha for bull highly volatile and bear market while appears unstable, but rarely negative alpha on bull non-volatile market.

### 4.3 Possible Experimental Problems

The smaller the chosen period, the larger the market trend discrepancy. Because of the noisiness of the price signal, short periods may represent quite different market trend so the agent set tuned on previous period signal may poorly behave on the next one.

Hummingbot in its current implementation has poor control over base asset spent which may cause larger earnings on bear market, compared to the competing agent families.

NIOX agent was used with irregular grid skewed spreads which may be the cause of a good performance only for bear market.

## 5 Further Improvements

Given the experiments that we have run, and experience gained while developing the infrastructure for the experiments, the following improvements can be considered.

A denser regular bid/ask spread grid may be implemented for more precise strategy selection. In the above-mentioned experiments, grid density was limited by the available computational resources while better AMSA performance could be expected with a more precisely tuned bid/ask grid with more fine-grained levels.

Hanging orders within the period may be involved in the experiments. Hanging orders were only used by Base agent setup but were disabled for both NIOX and Hummingbot configurations.

Hanging orders throughout the periods may be implemented. Long-lasting orders are not currently implemented so even if an agent is performing successfully in the previous period and re-selected for the next period, it has all of its orders canceled at the period boundary. Keeping the orders hanging across the periods may improve the return for winning agents and increase the overall performance.

Base/quote order amount grid might be finer grained as maximum available inventories were used in the current setup, while in some circumstances smaller bid/ask orders may improve the overall return.

Agent selection policy tuning may be improved. As of now, agents achieving positive return and alpha during the previous interval backtesting and real trading are selected for the next round of trades in the current AMSA version. More sophisticated selection algorithm may improve the overall return.

As an extension or variation of the above while running the AMSA in real trading environments, front testing (also called "paper trading") can be performed on live market data instead of backtesting on historical data. This could be done following the same simulation of the LOB execution as we have described but might be more realistic being run on live data in sync with the real market making.

The inventory funds distribution policy might be changed to uneven (prioritized or weighted) among the agents involved in real market making, giving more funds to more successful agents can be explored and one of the measures as it might increase the overall returns because of greater contribution of more successful strategies.

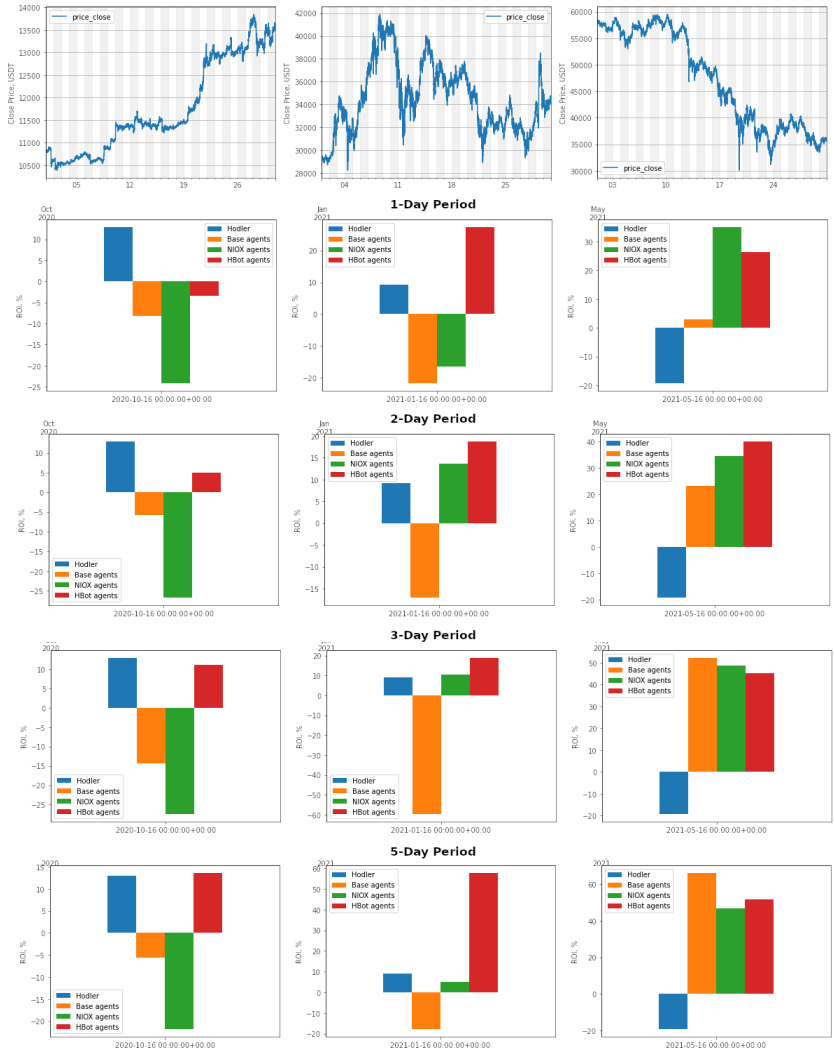

**Fig. 1.** Results (ROI %) of minute-based simulation (backtesting) of three agent families compared to "Hodler" across three different market types with different durations of evaluation periods.

## 6  Conclusion and Future Work

We have presented the architecture of adaptive multi-strategy agent (AMSA) for autonomous market making as a Narrow AGI solution applied to financial domain. The architecture has been evaluated in a market making simulation framework by means of backtesting of the limit order book operations relying on full scope of historical market data for BTC/USDT trading pair on Binance crypto exchange during three months with different market conditions with different time granularities.

The evaluation has been applied to three different families of market making agents. One of the families (namely, open-source Hummingbot implementation) was found to be capable of providing both non-negative return and "alpha" (excess return over conservative "hodling" strategy) across all evaluated market conditions.

Our further work will be dedicated to exploring the applicability of our market making AMSA architecture for real market making relying on Hummingbot and extending our studies on other trading pairs and exchanges.